\documentclass[conference]{IEEEtran}
\usepackage{amsmath}
\usepackage{amsthm}
\usepackage{amssymb}
\usepackage{bm}
\usepackage{xspace}
\usepackage{xcolor}
\usepackage{graphicx}
\usepackage{psfrag}
\usepackage{url}
\usepackage{framed}
\usepackage{cite}
\usepackage{algorithm}
\usepackage{algorithmicx}
\usepackage{algpseudocode}
\usepackage{float}

\theoremstyle{plain}

\newtheorem{proposition}{Proposition}

\theoremstyle{definition}

\theoremstyle{plain}

\theoremstyle{definition}

\theoremstyle{remark}

\newcommand{\de}{\,\mathrm{d}}
\newcommand{\vecp}{\mathbf{p}}
\newcommand{\vecw}{\mathbf{w}}
\newcommand{\vecq}{q}
\newcommand{\vecx}{\mathbf{x}}
\newcommand{\entop}{\mathcal{H}}

\DeclareMathOperator{\kl}{D}

\DeclareMathOperator*{\argmin}{argmin}

\title{Capacity Achieving Modulation for Fixed Constellations with Average Power Constraint}

\IEEEoverridecommandlockouts

\author{\IEEEauthorblockN{Georg B\"ocherer, Fabian Altenbach, and Rudolf Mathar}
\IEEEauthorblockA{Institute for Theoretical Information
Technology\\
RWTH Aachen University, 52056 Aachen, Germany
\\ Email: \texttt{\{boecherer,altenbach,mathar\}@ti.rwth-aachen.de}}
\thanks{This work has been supported by the UMIC Research Center, RWTH
Aachen University.}
}

\begin{document}
\maketitle

\begin{abstract}
The capacity achieving probability mass function (PMF) of a finite signal constellation with an average power constraint is in most cases non-uniform. A common approach to generate non-uniform input PMFs is Huffman shaping, which consists of first approximating the capacity achieving PMF by a sampled Gaussian density and then to calculate the Huffman code of the sampled Gaussian density. The Huffman code is then used as a prefix-free modulation code. This approach showed good results in practice, can however lead to a significant gap to capacity. In this work, a method is proposed that efficiently constructs optimal prefix-free modulation codes for any finite signal constellation with average power constraint in additive noise. The proposed codes operate as close to capacity as desired. The major part of this work elaborates an analytical proof of this property. The proposed method is applied to 64-QAM in AWGN and numeric results are given, which show that, opposed to Huffman shaping, by using the proposed method, it is possible to operate very close to capacity over the whole range of parameters.
\end{abstract}

\section{Introduction}

Reliable communication over a noisy channel at maximum rate is only possible if the input is distributed according to a capacity achieving distribution, i.e., a distribution for which the mutual information between channel input and channel output is maximum.

In digital communication systems, the input is not continuous but has to be chosen from a discrete and finite constellation of signal points. In addition, a modulator has to generate the probability mass function (PMF) of the signal points from equiprobable binary input data. The idea to do so by prefix-free modulation codes originates in \cite[IV.A]{Forney1984}. Based on this idea, Huffman Shaping was developed in \cite{Kschischang1993,Ungerboeck2002,Ungerboeck2008}. Huffman Shaping consists of two steps. First, the PMF of the signal points that minimizes the average energy subject to fixed entropy is chosen. The solution of this optimization problem is a sampled Gaussian density \cite{Kschischang1993},\cite[Sec. 4.1.2]{Fischer2002}. An equivalent formulation of this approach is to look for the signal point PMF that maximizes entropy subject to an average power constraint. Then, in a second step, the Huffman code of the obtained sampled Gaussian density is used as a prefix-free modulation code. However, Huffman Shaping is sub-optimal and can lead to non-trivial gaps to capacity \cite[Sec. VIII.A]{Wachsmann1999},\cite[Sec. 4.2.6]{Fischer2002}. The reason is that maximizing input entropy is in general not equivalent to maximizing the mutual information between input and output. Furthermore, the distance measure minimized by Huffman coding is not appropriate \cite{Bocherer2011}.

In this work, we propose a method to derive optimal prefix-free modulation codes for fixed signal constellations with an average power constraint and additive noise. We first show that for every fixed signal constellation, average power constraint, and additive noise density, the capacity achieving PMF is given by the solution of a convex optimization problem that can efficiently be solved numerically. We then use Geometric Huffman Coding \cite{Bocherer2011} to find prefix-free modulation codes that approximate capacity achieving PMFs. We finally prove that our method approximates any capacity achieving PMF arbitrarily well both with respect to (w.r.t.) the resulting average power and mutual information. As an illustration of our results, we apply our method to 64-QAM in AWGN and observe that capacity can be approximated extremely well with prefix-free modulation over the whole range of the average power constraint. Our method differs from Huffman Shaping in two ways: first, we approximate the capacity achieving PMFs, which are very different from the sampled Gaussian density over a large range of the average power constraint. Second, the capacity achieving prefix-free modulation codes are not Huffman.

The remainder of this work is organized as follows. In Section~\ref{sec:problem} we state the problem of finding good prefix-free modulation codes. Capacity achieving PMFs are characterized in Section~\ref{sec:optimization}. We then derive in Section~\ref{sec:approximation} the offset in mutual information that results from using a `wrong' PMF. In Section~\ref{sec:coding}, we show how to find optimal prefix-free modulation codes. Finally, we present numerical results for 64-QAM in Section~\ref{sec:numeric}.

\section{Problem Statement}
\label{sec:problem}
Consider the discrete-time memoryless channel with additive noise given by $Y = X + Z$.
The input $X$ takes values in the finite signal constellation set $\mathcal{X}\!\subset\!\mathbb{C}$ of cardinality $|\mathcal{X}|\!=\!m$. Input $X$ is subject to an average power constraint $\bar{E}$ in terms of energy per channel use. The additive noise term $Z$ takes values in $\mathbb{C}$ and is distributed according to a density $h$. For each $i=1,\dotsc,m$, conditioned on $x_i\in\mathcal{X}$, the channel output $Y$ is distributed according to $h_i(y)\triangleq h(y-x_i)$.

Let the input $X$ be independent identically distributed (IID) according to a PMF $\mathbf{p}=(p_1,\dotsc,p_m)^T$. We denote the energy of the $i$th signal point by $w_i=|x_i|^2$ and define the energy vector as $\mathbf{w}=(w_1,\dotsc,w_m)^T$. The average power constraint $\bar{E}$ can now be written as $\vecw^T \vecp \leq \bar{E}$. A capacity achieving PMF maximizes the mutual information between $X$ and $Y$. According to \text{\cite[Eq. (2.4.40)]{Gallager1968}}, the mutual information is given by
\begin{align}
\mathcal{I}(\mathbf{p}) & = \entop(Y) - \entop(Y|X) \\ 
& = - \int_\mathbb{C} \Bigl(\sum_i p_i h_i(y)\Bigr) \log\Bigl( \sum_i p_i h_i(y) \Bigr) \de y \nonumber \\ &\quad +  \sum_i p_i \int_\mathbb{C} h_i(y) \log[h_i(y)] \de y \label{eq:mutualInformation}
\end{align}
where $\entop(\cdot)$ denotes the entropy function.

In prefix-free modulation, the binary data stream \text{$(B_k,k\in\mathbb{N})$} is parsed into words from a full prefix-free code, and each word is mapped to a signal point from $\mathcal{X}$ by a one-to-one mapping, see, e.g., \cite{Abrahams1998}. The data bits $B_k$ are equiprobable and jointly independent. As a consequence, the parsed words are independent and identically distributed according to $p_i=2^{-\ell_i}$, i.e., the probability to parse word $i$ is $2^{-\ell_i}$, where $\ell_i$ is the number of bits in the $i$th word. We define the set $\mathbb{D}$ of \emph{dyadic PMFs} as
\begin{align}
\mathbb{D} = \{\mathbf{p}\,|\,\mathbf{p}&\text{ is a PMF},\nonumber\\&\text{for } i=1,\dotsc,m : p_i=2^{-\ell_i}, \ell_i\in\mathbb{N}\}.
\end{align}
By the Kraft inequality \cite[Theorem~5.2.1]{Cover2006}, any $\mathbf{p}\in\mathbb{D}$ can be generated by parsing $(B_k)$ by a prefix-free code with the corresponding codeword lengths. Searching for good prefix-free modulation codes is thus equivalent to searching for good dyadic input PMFs. The rest of this work is therefore about dyadic input PMFs, but keep in mind that for each dyadic input PMF, appropriate prefix-free modulation codes are at hand.

Restricting the input PMF $\vecp$ to the set of dyadic PMFs $\mathbb{D}$, the discrete optimization problem becomes
\begin{align}
&\text{maximize } && \mathcal{I}(\vecp)\nonumber\\
&\text{subject to } &&  p_i = 2^{-\ell_i}, \ell_i \in \mathbb{N}, i = 1,\ldots,m\nonumber\\
 		  &&& \mathbf{1}^T\vecp=1,\quad\mathbf{w}^T \vecp \leq \bar{E} \label{eq:discreteProblem}
\end{align}
where $\mathbf{1}$ is a $m\times 1$ vector with all components equal to one. This is a non-linear optimization problem with integer constraint and no efficient algorithm to solve it is known. Our approach is therefore to first drop the integer constraint and to calculate capacity achieving (in general non-dyadic) PMFs and then to approximate these capacity achieving PMFs by dyadic PMFs. We will do so in the following sections.

\section{Capacity Achieving PMFs}
\label{sec:optimization}
Mutual information is concave in $\vecp$, which can be seen as follows. The first term in \eqref{eq:mutualInformation} is an integral over a positively weighted sum of functions concave in $p_i$, and is thereby concave \cite[Ch. 3.2]{Boyd2004}. The second term is a linear function in $\vecp$. Therefore, the mutual information is concave in $\mathbf{p}$. For convenience, we replace the maximization in \eqref{eq:mutualInformation} by minimization. This leads to the optimization problem
\begin{equation}
\begin{array}{ll}
\text{minimize } & -\mathcal{I}(\mathbf{p}) \\
\text{subject to } &\mathbf{1}^T \mathbf{p} = 1,\quad\mathbf{p} \geq 0 \\
 		  & \mathbf{w}^T \mathbf{p} \leq \bar{E}.\label{eq:continuousProblem}
 \end{array}
\end{equation}
Since the objective function and the inequality constraints are convex and the equality constraint is affine, the above optimization problem is convex. Therefore, an optimal solution can be calculated efficiently by numerical optimization methods \cite{Boyd2004}. We now explicitly evaluate the Karush-Kuhn-Tucker (KKT) conditions for \eqref{eq:continuousProblem}. We refer to these conditions later in this work. The Lagrangian of the optimization problem \eqref{eq:continuousProblem} is given by 
\begin{align}
 L(\mathbf{p},\pmb{\mu},\nu,\lambda) & = -\mathcal{I}(\mathbf{p}) - \pmb{\mu}^T \mathbf{p} + \nu(\mathbf{w}^T \mathbf{p} - \bar{E})\nonumber\\
& \qquad+ \lambda (\mathbf{1}^T\mathbf{p} - 1)
\end{align}
with dual variables $\pmb{\mu} \in \mathbb{R}^m$ and $\nu,\lambda \in \mathbb{R}$. Assuming primal feasibility, for each $1\leq i \leq m$, the KKT conditions are
\begin{align}
 \frac{\partial L(\mathbf{p},\pmb{\mu},\nu,\lambda)}{\partial p_i} & = - \frac{\partial \mathcal{I}(\mathbf{p})}{\partial p_i} - \mu_i + \nu w_i + \lambda = 0 \label{eq:derivativeLagrangian}\\
& \quad\, \mu_ip_i = 0,\quad \nu (\mathbf{w}^T\mathbf{p} - \bar{E}) = 0 \label{eq:complemslack}\\
& \quad\,  \mu_i \geq 0, \quad \nu \geq 0. \label{eq:dualfeas}
\end{align}
Denote by $\vecp^*,\pmb{\mu}^*,\lambda^*,\nu^*$ a tuple that fulfills the KKT conditions. From dual feasibility \eqref{eq:dualfeas} it follows that $\mu^*_i \geq 0$. For every $p^*_i > 0$, by complementary slackness \eqref{eq:complemslack}, we have $\mu^*_i = 0$. Using these observations in \eqref{eq:derivativeLagrangian} and rearranging the terms, we get
\begin{align}
\frac{\partial \mathcal{I}(\mathbf{p^*})}{\partial p^*_i} & \leq \lambda^* + \nu^* w_i,\quad \text{with equality if } p^*_i > 0.\label{eq:kkt}
\end{align}

\section{Using the `Wrong' PMF}
\label{sec:approximation}
\begin{figure}
\centering
\def\svgwidth{0.9\columnwidth}
\footnotesize
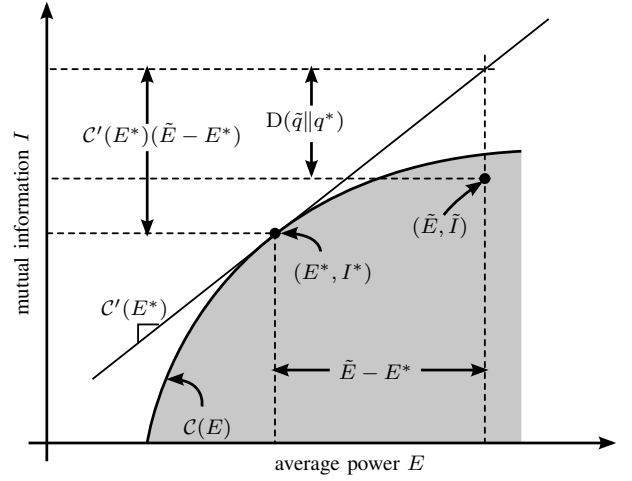
\caption{The shaded area below the capacity curve $\mathcal{C}(E)$ is the region of valid operating points $\{(\mathbf{w}^T\mathbf{p},\mathcal{I}(\mathbf{p}))\,|\,\mathbf{p}\text{ is an input PMF}\}$. Consider a capacity achieving PMF $\vecp^*$ and some PMF $\tilde{\vecp}$. If $\tilde{p}_i=0$ whenever $p_i^*=0$, then the corresponding operating points $(E^*,I^*)$ and $(\tilde{E},\tilde{I})$ relate as $
\tilde{I} = I^*+\mathcal{C}'(E^*)(\tilde{E}-E^*)-\kl(\tilde{q}\Vert q^*)$ where $\tilde{q}$ and $q^*$ are the output densities that correspond to $\tilde{\vecp}$ and $\vecp^*$, respectively.
\vspace{-0.55cm}
}
\label{fig:geometry}
\end{figure}
Suppose our average power constraint is $E^*$. Solving \eqref{eq:continuousProblem} for $\bar{E}=E^*$ yields the corresponding capacity achieving PMF $\vecp^*$ and the mutual information $I^*=\mathcal{I}(\vecp^*)$. Our target operating point is thus $(E^*,I^*)$. If we use some other PMF $\tilde{\vecp}$ as an approximation of $\vecp^*$, the effective operating point is $(\tilde{E},\tilde{I})$, where $\tilde{E}=\vecw^T\tilde{\vecp}$ and $\tilde{I}=\mathcal{I}(\tilde{\vecp})$. We denote by $\mathcal{C}(E)$ the capacity curve, i.e., the maximum mutual information that is achievable under the average power constraint $\bar{E}=E$. Formally,
\begin{align}
\mathcal{C}(E) = \mathcal{I}(\vecp):\vecp \text{ is a solution of \eqref{eq:continuousProblem} for } \bar{E} = E.
\end{align}
Note that $\mathcal{I}(\vecp^*)=\mathcal{C}(E^*)$ but in general $\mathcal{I}(\tilde{\vecp})\neq\mathcal{C}(\tilde{E})$.
 The following proposition shows how the effective operating point $(\tilde{E},\tilde{I})$ relates to the target operating point $(E^*,I^*)$ in terms of $\tilde{\vecp}$ and $\vecp^*$. A visualization of the proposition is given in Fig.~\ref{fig:geometry}.
\begin{proposition}
\label{prop:geometry}
Consider a capacity achieving PMF $\vecp^*$ and some input PMF $\tilde{\vecp}$. If 
\begin{align}
\tilde{p}_i=0 \text{ whenever } p_i^*=0\label{eq:condition}
\end{align}
then the corresponding operating points $(\tilde{E},\tilde{I})$ and $(E^*,I^*)$ relate as follows:
\begin{align}
\tilde{E} &= \vecw^T\tilde{\vecp},\qquad E^*=\vecw^T\vecp^*\\
\tilde{I} &= I^*+\mathcal{C}'(E^*)(\tilde{E}-E^*)-\kl(\tilde{\vecq}\Vert \vecq^*)\label{eq:geometry}
\end{align}
where $\tilde{\vecq}$ and $\vecq^*$ are the output densities that correspond to $\tilde{\vecp}$ and $\vecp^*$, respectively and where $\mathcal{C}'(E)$ is the derivative of $\mathcal{C}(E)$ w.r.t. $E$. $\kl(\cdot\Vert\cdot)$ denotes the Kullback-Leibler (KL) distance as defined in \cite[Sec.~8.5]{Cover2006}.
\end{proposition}
\begin{IEEEproof}
The output density $q$ that results from using the input PMF $\vecp$ is given by
\begin{align}
q(y)=\sum_i p_i h_i(y),\qquad y\in\mathbb{C}.
\end{align}
Denote by $\tilde{q}$ and $q^*$ the output densities that result from using the input PMF $\tilde{\vecp}$ and $\vecp^*$, respectively. We now have
\begin{align}
\tilde{I}=\mathcal{I}(\tilde{\vecp})&=\sum_i \tilde{p}_i\int_\mathbb{C} h_i(y)\log\frac{h_i(y)}{\tilde{q}(y)}\de y\\
&=\sum_i \tilde{p}_i\int_\mathbb{C} h_i(y)\log\frac{h_i(y)q^*(y)}{\tilde{q}(y)q^*(y)}\de y\\
&=\sum_x \tilde{p}_i\int_\mathbb{C} h_i(y)\log\frac{h_i(y)}{q^*(y)}\de y
\nonumber
\\&\qquad\qquad+\sum_i \tilde{p}_i\int_\mathbb{C} h_i(y)\log\frac{q^*(y)}{\tilde{q}(y)}\de y.\label{eq:integralTerm}
\end{align}
For the second summand in \eqref{eq:integralTerm}, we further get
\begin{align}
\sum_i \tilde{p}_i\int_\mathbb{C} &h_i(y)\log\frac{q^*(y)}{\tilde{q}(y)}\de y=\int_\mathbb{C} \Bigl(\sum_i \tilde{p}_i h_i(y)\Bigr)\log\frac{q^*(y)}{\tilde{q}(y)}\de y\nonumber\\
&=\int_\mathbb{C} \tilde{q}(y)\log\frac{q^*(y)}{\tilde{q}(y)}\de y
=-\kl(\tilde{\vecq}\Vert \vecq^*).\label{eq:klTerm}
\end{align}
By simple calculus, we get in accordance with \cite[Eq. (4.5.5)]{Gallager1968} for the partial derivatives of $\mathcal{I}(\vecp)$ with respect to $p_i$
\begin{align}
\frac{\partial\mathcal{I}(\vecp)}{\partial p_i} = \int_\mathbb{C} h_i(y)\log\frac{h_i(y)}{q(y)}\de y- \log e\label{eq:partialDerivative}
\end{align}
Using \eqref{eq:kkt} and \eqref{eq:partialDerivative}, we get for the integral term in the first summand in \eqref{eq:integralTerm}
\begin{align}
\int_\mathbb{C} h_i(y)\log\frac{h_i(y)}{q^*(y)}\de y&=\frac{\partial\mathcal{I}(\vecp^*)}{\partial p^*_i}+\log e\\
&\leq \lambda^*+\nu^* w_i +\log e\label{eq:inequality}
\end{align}
with equality if $p^*_i > 0$. Expectation w.r.t. $\tilde{\vecp}$ gives
\begin{align}
&\sum_i \tilde{p}_i\int_\mathbb{C}h_i(y)\log\frac{h_i(y)}{q^*(y)}\de y=
\sum_i \tilde{p}_i(\lambda^*+\nu^* w_i+\log e)\label{eq:pzero}\\
&=\lambda^* + \nu^* \sum_i \tilde{p}_i w_i + \log e\\
&=\lambda^* + \nu^* \sum_i (\tilde{p}_i-p^*_i+p^*_i) w_i +\log e\\
&=\lambda^* + \nu^* \sum_i p^*_i w_i +\log e+\nu^*\sum_i (\tilde{p}_i-p^*_i) w_i\\
&=\sum_i p^*_i(\lambda^* + \nu^*  w_i + \log e) + \nu^*(\tilde{E}-E^*)\label{eq:powerTerms}\\
&=\sum_i p_i^*\int_\mathbb{C}h_i(y)\log\frac{h_i(y)}{q^*(y)}\de y+ \nu^*(\tilde{E}-E^*)\\
&=I^*+\nu^*(\tilde{E}-E^*)\label{eq:energyTerm}
\end{align}
where we have equality in \eqref{eq:pzero} since, according to the assumption of the proposition, $\tilde{p}_i=0$ whenever $p^*_i=0$, and for all $i$ with $p_i^*>0$, we have because of \eqref{eq:kkt} equality in \eqref{eq:inequality}. In \eqref{eq:powerTerms}, we used $\tilde{E}=\vecw^T\tilde{\vecp}$ and $E^*=\vecw^T\vecp^*$, respectively. Using \eqref{eq:energyTerm} and \eqref{eq:klTerm} in \eqref{eq:integralTerm}, we get
\begin{align}
\tilde{I} = I^*+\nu^*(\tilde{E}-E^*)-\kl(\tilde{\vecq}\Vert \vecq^*).\label{eq:withNu}
\end{align}
By \cite[Ch.~5.6.3]{Boyd2004} it holds that
\begin{align}
\nu^* = \left.\frac{\partial\mathcal{C}(E)}{\partial E}\right\vert_{E=E^*} \triangleq \mathcal{C}'(E^*).\label{eq:nu}
\end{align}
Using \eqref{eq:nu} in \eqref{eq:withNu} gives the statement of the proposition.
\end{IEEEproof}
%
%
%
%

Proposition~\ref{prop:geometry} is formulated for the case where consecutive signal points are generated independently according to the same PMF $\vecp$. We now consider the case where $n$ consecutive signal points are generated according to a joint PMF $\vecp^{(n)}$. The resulting average power and mutual information per block become respectively
\begin{align}
E^{(n)}=\mathcal{E}(\vecp^{(n)}),\qquad I^{(n)} = \mathcal{I}(\vecp^{(n)})
\end{align}
where $\mathcal{E}(\vecp^{(n)})$ is defined as
\begin{align}
\mathcal{E}(\vecp^{(n)}) = \sum_{\vecx\in\mathcal{X}^n}p^{(n)}(\vecx)\Vert\vecx\Vert^2
\end{align}
where $\mathcal{X}^n$ denotes the Cartesian product of $n$ copies of $\mathcal{X}$ and where $\Vert\cdot\Vert$ denotes the Euclidean norm.
Since the channel is memoryless, a capacity achieving joint PMF $\vecp^{(n)*}$ is the product of $n$ copies of some $\vecp^*$. As a consequence, $I^{(n)*}=nI^*$ and $E^{(n)*}=nE^*$. The capacity curve is $\mathcal{C}^{(n)}(E^{(n)*})=n\mathcal{C}(E^{(n)*}/n)$. Consequently, we get for the derivative
\begin{align}
\frac{\partial \mathcal{C}^{(n)}(E^{(n)*})}{\partial E^{(n)*}}&=\frac{\partial n\mathcal{C}(\frac{E^{(n)*}}{n})}{\partial E^{(n)*}}\\
&=n\cdot \mathcal{C}'\Bigl(\frac{E^{(n)*}}{n}\Bigr)\cdot\frac{1}{n}
=\mathcal{C}'(E^*).
\end{align}
For blocks of $n$ symbols, Proposition~\ref{prop:geometry} now becomes
\begin{align}
\tilde{I}^{(n)} = \tilde{I}^{(n)*}+\mathcal{C}'(E^*)(\tilde{E}^{(n)}-nE^*)-\kl(\tilde{\vecq}^{(n)}\Vert\vecq^{(n)*})
\end{align}
where $\tilde{\vecq}^{(n)}$ and $\vecq^{(n)*}$ are the output densities that result from using the input PMFs $\tilde{\vecp}^{(n)}$ and $\vecp^{(n)*}$, respectively.
Dividing by $n$ we get for the mutual information per channel use $\tilde{I}_n=\tilde{I}^{(n)}/n$ and the energy per channel use $\tilde{E}_n=\tilde{E}^{(n)}/n$
\begin{align}
\tilde{I}_n = I^*+\mathcal{C}'(E^*)(\tilde{E}_n-E^*)-\frac{\kl(\tilde{\vecq}^{(n)}\Vert\vecq^{(n)*})}{n}.\label{eq:blockOP}
\end{align}

\section{Optimal Dyadic PMFs}
\label{sec:coding}
For a given average power constraint $E^*$, we want to find an operating point $(\tilde{E},\tilde{I})$ that is close to the target operating point $(E^*,I^*=\mathcal{C}(E^*))$ both in terms of average power and mutual information. The following proposition gives sufficient conditions to accomplish this.
\begin{proposition}\label{prop:sufficientCondition}
Consider a capacity achieving PMF $\vecp^*$ and a sequence of PMFs $(\tilde{\vecp}_n, n\in\mathbb{N})$ where each PMF in the sequence fulfills condition \eqref{eq:condition}. Assume further that $\mathcal{C}(E)$ is strictly concave in $E$. Then
\begin{align}
(\tilde{E}_n,\tilde{I}_n)\overset{n\rightarrow\infty}{\longrightarrow}(E^*,I^*)
\end{align}
if one of the following two properties holds:
\begin{align}
\text{Property 1:}&\qquad\kl(\tilde{\vecq}_n\Vert\vecq_n^*)\overset{n\rightarrow\infty}{\longrightarrow} 0\\
\text{Property 2:}&\qquad\kl(\tilde{\vecp}_n\Vert\vecp_n^*)\overset{n\rightarrow\infty}{\longrightarrow} 0.
\end{align}
\end{proposition}
\begin{IEEEproof}
The assumption that $\mathcal{C}(E)$ is strictly concave in $E$ implies the sufficiency of Property 1. This can best be seen by considering the visualization of Proposition~\ref{prop:geometry} in Fig.~\ref{fig:geometry}. As $\kl(\tilde{\vecq}\Vert\vecq^*)$ becomes smaller, $(\tilde{E},\tilde{I})$ is approaching the tangent in $(E^*,I^*)$ of the boundary. However, because the tangent is linear and the boundary is strictly concave, as $\kl(\tilde{\vecq}\Vert\vecq^*)$ is getting smaller, $(\tilde{E},\tilde{I})$ has to walk in the direction of $(E^*,I^*)$. Otherwise, $(\tilde{E},\tilde{I})$ would go above the boundary, which is impossible since by the definition of the boundary, $\tilde{I}\leq\mathcal{C}(\tilde{E})$.

Sufficiency of Property 2 holds because the KL-distance between the output densities is upper bounded by the KL-distance between the input PMFs, i.e,
\begin{align}
\kl(\tilde{\vecq}\Vert\vecq^*)\leq\kl(\tilde{\vecp}\Vert\vecp^*)
\end{align}
This can easily be shown along the lines of \cite[Sec.~4.4]{Cover2006}. Thus, Property 2 implies Property 1, and the sufficiency of Property 1 was shown in the first part of this proof.
\end{IEEEproof}

We now come to the central point of this work, namely to approximate a target operating point $(E^*,I^*)$ by a dyadic PMF $\tilde{\vecp}\in\mathbb{D}$. By Proposition~\ref{prop:geometry}, we know that minimizing the KL-distance $\kl(\tilde{\vecq}\Vert\vecq^*)$ between the corresponding output densities maximizes mutual information in the sense that $(\tilde{E},\tilde{I})$ approaches the boundary $\mathcal{C}(E)$, and furthermore, by Proposition~\ref{prop:sufficientCondition}, we know that if $\kl(\tilde{\vecq}\Vert\vecq^*)$ approaches zero, then $(\tilde{E},\tilde{I})$ converges to $(E^*,I^*)$ both in terms of average power and mutual information. By Property~2 in Proposition~\ref{prop:sufficientCondition}, we know that both effects can also be achieved by minimizing $\kl(\tilde{\vecp}\Vert\vecp^*)$. No algorithm is known that finds the dyadic PMF that minimizes $\kl(\tilde{\vecq}\Vert\vecq^*)$ with complexity polynomial in $m$. We therefore minimize the KL-distance between the input PMFs, i.e., the aim is to solve 
\begin{align}
\tilde{\vecp}=\argmin_{\mathbf{p}\in\mathbb{D}}\kl(\mathbf{p}\Vert \mathbf{p}^*).
\end{align}
As shown in \cite{Bocherer2011}, $\tilde{\vecp}$ can efficiently be found by \emph{Geometric Huffman Coding} (\textsc{Ghc}). For the definition of \textsc{Ghc} and an implementation see \cite{Bocherer2011,website:ghc}. The complexity of \textsc{Ghc} is $m\log m$. In the following, $\tilde{\vecp}=\textsc{Ghc}(\vecp^*)$, i.e., $\tilde{\vecp}$ denotes the optimal dyadic approximation of the capacity achieving PMF $\vecp^*$. Because of the discrete nature of $\mathbb{D}$, $\kl(\tilde{\vecp}\Vert\mathbf{p}^*)>0$ in most cases, and as a consequence, there is a non-zero gap between the reached operating point $(\tilde{E},\tilde{I})$ and the target operating point $(E^*,I^*)$. This gap can be made arbitrarily small by using \textsc{Ghc} to approximate the capacity achieving joint PMF $\mathbf{p}^{(n)*}$ of $n$ consecutive signal points. Since $\tilde{\vecp}^{(n)}=\textsc{Ghc}(\vecp^{(n)*})$, by \cite[Proposition 2]{Bocherer2011}, we have
\begin{align}
\frac{\kl(\tilde{\vecp}^{(n)}\Vert \vecp^{(n)*})}{n}\overset{n\rightarrow\infty}{\longrightarrow}0
\end{align}
Plugging this into \eqref{eq:blockOP}, we get by the same concavity argument as in the proof of Proposition~\ref{prop:sufficientCondition} the following.
\begin{proposition}\label{prop:mainResult}
Consider a capacity achieving PMF $\vecp^*$ and the corresponding capacity achieving joint PMFs $\vecp^{(n)*}$. Assume $\mathcal{C}(E)$ is strictly concave in $E$. For 
\begin{align}
\tilde{\vecp}^{(n)}&\triangleq\textsc{Ghc}(\vecp^{(n)*})\\
\text{and }(\tilde{E}_n,\tilde{I}_n)&\triangleq\Bigr(\frac{\mathcal{E}(\tilde{\vecp}^{(n)})}{n},\frac{\mathcal{I}(\tilde{\vecp}^{(n)})}{n}\Bigr),
\end{align}
we have
\begin{align}
(\tilde{E}_n,\tilde{I}_n)\overset{n\rightarrow\infty}{\longrightarrow}(E^*,I^*).
\end{align}
\end{proposition}

\section{Numerical Results: $64$-QAM}
\label{sec:numeric}

\begin{figure*}
\parbox[t]{\textwidth}{
\def\svgwidth{\textwidth}
\footnotesize
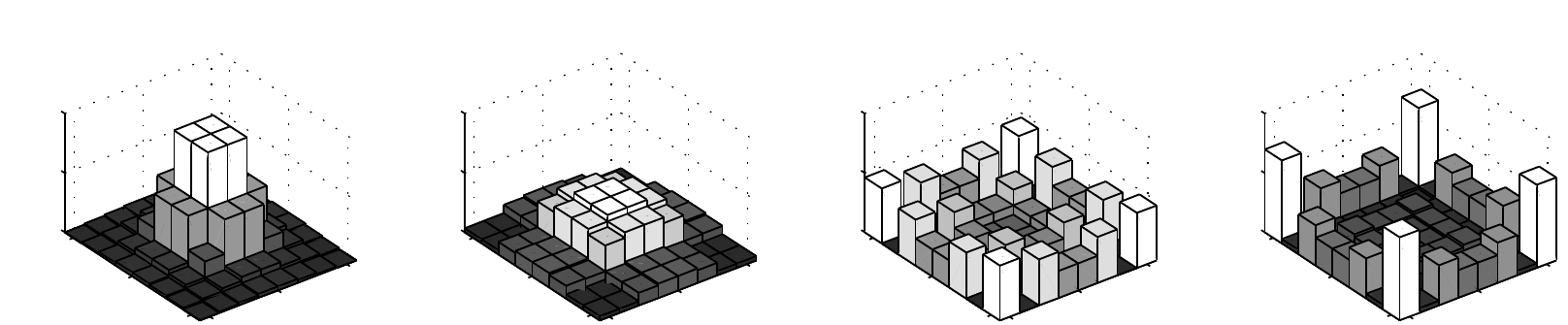
\caption{Capacity achieving PMFs for the $8\times 8$ signal points of 64-QAM for the average power constraints $\bar{E}=2.5,5,10,20$. The probabilities of the signal points are given by the heights of the vertical bars.}
\label{fig:mountains}
}\\[0.2cm]
\parbox[t]{0.32\textwidth}{
\def\svgwidth{0.32\textwidth}
\footnotesize
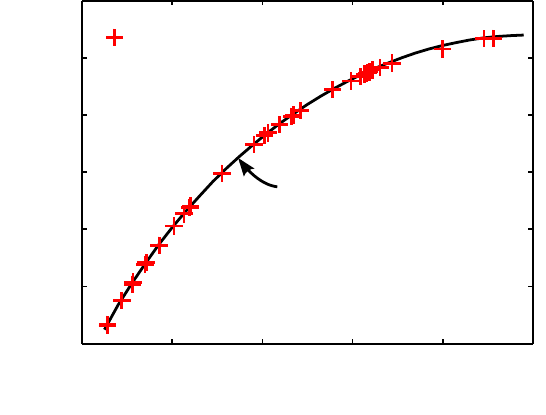
\caption{}
\label{fig:operating_points}
}
\hfill
\parbox[t]{0.32\textwidth}{
\def\svgwidth{0.32\textwidth}
\footnotesize
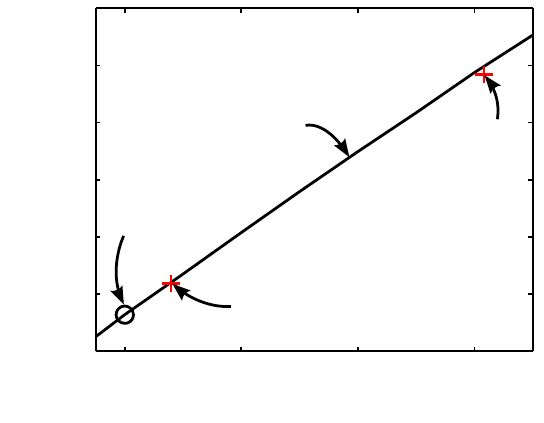
\caption{}
\label{fig:convergence}
}
\hfill
\parbox[t]{0.32\textwidth}{
\def\svgwidth{0.32\textwidth}
\footnotesize
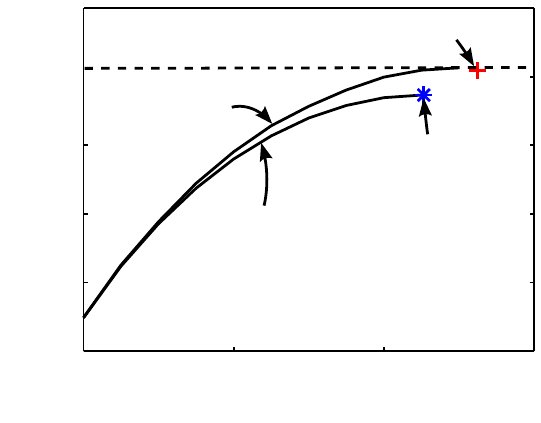
\caption{}
\label{fig:sampled_gaussian}
}
\vspace{-0.5cm}
\end{figure*}
For illustrative purpose, we apply our algorithm to 64-QAM. The additive noise is zero-mean circular symmetric white Gaussian of unit variance. The scaling of 64-QAM is specified through the highest signal point energy $\max|x|^2$.

For $\max|x|^2=20$, and for the average power constraints of $\bar{E}=2.5,5,10,20$, the capacity achieving PMFs are displayed in Fig.~\ref{fig:mountains}. The PMFs are obtained by solving \eqref{eq:continuousProblem}. For $\bar{E}=2.5,5$, the PMFs resemble the sampled Gaussian density, but for $\bar{E}=10$, the signal point probabilities follow no longer a monotonic function of the signal point energy. For $\bar{E}=20$, the average power constraint is no longer active and the resulting average power of the capacity achieving PMF is $E=11.91$.

We now calculate the dyadic approximations of the capacity achieving operating points for $\bar{E}=2.5, 2.6, 2.7,\dotsc,12$. For $\bar{E}>11.91$, the average power constraint is no longer active. In Fig.~\ref{fig:operating_points}, the dyadic operating points are displayed. The dyadic operating points are very close to the capacity curve $\mathcal{C}(E)$. This illustrates that minimizing $\kl(\vecp\Vert\vecp^*)$ gives good results in practice. However, the placement of dyadic operating points is irregular. So, for some optimal operating points, there is no close dyadic operating point. This problem is discussed next.

We now illustrate how a specific (capacity achieving) target operating point can be approximated closely by block modulation. The results are displayed in Fig.~\ref{fig:convergence}. We choose as a target operating point $(E^*,I^*)=(5.20,1.81)$. Denote the corresponding capacity achieving PMF by $\vecp^*$. Using the dyadic PMF $\tilde{\vecp}=\text{\textsc{Ghc}}(\vecp^*)$, the resulting dyadic operating point is $(\tilde{E}_1,\tilde{I}_1)=(5.82,1.90)$, which corresponds to  approximation errors of $10.65\%$ and $4.74\%$ w.r.t. average power and mutual information, respectively. Exceeding the power constraint by more than $10\%$ may be critical. Jointly modulating two consecutive signal points, i.e., using the joint dyadic PMF $\tilde{\vecp}^{(2)}=\text{\textsc{Ghc}}(\vecp^{(2)*})$ results in the dyadic operating point $(\tilde{E}_2,\tilde{I}_2)=(5.28,1.82)$, which corresponds to a power exceed of $1.52\%$ and $0.55\%$ of mutual information. This is a significant improvement and in accordance with Proposition~\ref{prop:mainResult}.

For $\max|x|^2=10$, we display in Fig.~\ref{fig:sampled_gaussian} the capacity curve $\mathcal{C}(E)$ and the mutual information curve $\mathcal{I}_\mathrm{SG}(E)$ that results from using sampled Gaussian densities for the input PMFs. For small $E$, both curves lie close together. However, as $E$ increases, there is an increasing gap. The capacity curve $\mathcal{C}(E)$ reaches its maximum $I=1.83$ for $E=6.98$. For $E>6.98$, the average power constraint is no longer active. Thus, $\mathcal{C}(\infty)=1.83$ is the capacity of the signal constellation without average power constraint. The curve $\mathcal{I}_\mathrm{SG}(E)$ reaches its maximum for $E=6.50$. The corresponding mutual information differs from the capacity $\mathcal{C}(\infty)$ by $-4.55\%$. While this gap is rather small, it implies a bottleneck when the aim is to communicate at rates very close to capacity. Also, this gap may be much larger for other signal constellations and/or other noise densities. The dyadic approximation of $\mathcal{C}(6.98)$ is within $-0.39\%$ of capacity $\mathcal{C}(\infty)$, while the dyadic approximation of $\mathcal{I}_\mathrm{SG}(6.50)$ is within $-4.52\%$ of capacity $\mathcal{C}(\infty)$.



%
%
%
%

\bibliographystyle{IEEEtran}
\normalsize
\bibliography{IEEEabrv,confs-jrnls,Dnc}

\end{document}